\documentclass[11pt]{amsart}
\usepackage{latexsym,amssymb,amsmath,amscd,amsthm}
\numberwithin{equation}{section}
\theoremstyle{remark}

\newtheorem{proposition}{{\bf PROPOSITION}}[section]
\newcommand{\bq}{\begin{equation}}
\newcommand{\bea}{\begin{array}}
\newcommand{\eea}{\end{array}}

\newcommand{\ga}{\alpha}
\newcommand{\gep}{\epsilon}
\newcommand{\gD}{\Delta}
\newcommand{\gl}{\lambda}
\newcommand{\gL}{\Lambda}
\newcommand{\gb}{\beta}

\newcommand{\mf}{\mathfrak}
\newcommand{\mc}{\mathcal}

\newcommand{\go}{\omega}
\newcommand{\gO}{\Omega}
\newcommand{\gG}{\Gamma}
\newcommand{\gt}{\theta}
\newcommand{\gs}{\sigma}

\newcommand{\gag}{\gamma}
\newcommand{\gd}{\delta}
\newcommand{\pp}{\partial}

\newcommand{\tl}{\tilde}
\newcommand{\na}{\nabla}
\newcommand{\gk}{\kappa}

\newcommand{\bs}{\blacksquare}

\newcommand{\bgs}{\bigstar}

\newcommand{{\DDD}}{D\!\!\!\!\!\!-}
\newcommand{\bx}{\Box}

\title{REMARKS ON THE FRIEDMAN EQUATIONS}

\author{Robert Carroll\\University of Illinois, Urbana, IL 61801}

\date{December, 2007\thanks{email: rcarroll@math.uiuc.edu}}

\begin{document}

\bibliographystyle{plain}

\begin{abstract} 
We give some heuristic results for FRW situations with Ricci flow.
\end{abstract}

\maketitle

%\tableofcontents

\section{INTRODUCTION}
\renewcommand{\theequation}{1.\arabic{equation}}
\setcounter{equation}{0}

In \cite{car,cczz,cari} we indicated some connections 
between geometry, quantum mechanics, Ricci flow, and gravity.  In \cite{cczz,cari} we began with the Perelman entropy functional
$({\bf 1A})\,\,{\mf F}=\int_M(R+|\na f|^2)exp(-f)dV$ and recalled that ${\mf F}$ is in fact
a Fisher information (following \cite{topp}).  One shows in \cite{muhl,topp} that the $L^2$
gradient flow of ${\mf F}$ is determined by evolution equations
\bq\label{1.1}
\pp_tg_{ij}=-2(R_{ij}+\na_i\na_jf);\,\,\pp_tf=-\gD f-R
\end{equation}
and this is equivalent to the decoupled family
\bq\label{1.2}
\pp_tg_{ij}=-2R_{ij};\,\,\pp_tf=-\gD f+|\na f|^2-R
\end{equation}
via a time dependent diffeomorphism (cf. also \cite{chln,khol,perl}).
We note from \cite{topp} that ${\mf F}$ is invariant under such diffeomorphisms and hence
we could deal with (1.1) as a generalized Ricci flow if desired.  Note the equations involving
$\psi$ are independent of choosing (1.1) or (1.2) since ${\mf F}$ is invariant and this
accounts for using a Ricci flow (and ({\bf 1C})) to obtain $P_t=-\gD P-RP$ leading to
$(\bgs)\,\,(1/m)div(P\na S)=\gD P-RP$.
By developments in \cite{ccz,cczz}
going back to Santamato's discussion of the the Schr\"odinger equation (SE) in a Weyl
space (cf. \cite{ccz,cczz,carl,cast,cama,sa}) and using the exact uncertainty principle of Hall-Reginatto (cf. \cite{ccz,cczz,carr,hall}) we could envision a Schr\"odinger wave
function $\psi=|\psi|exp(iS/\hbar)$ with $P=|\psi|^2\sim exp(-f)$ a probability density and
$({\bf 1B})\,\,Q=-(\hbar^2/16m)[R+(8/\sqrt{P})\gD\sqrt{P}]$ a quantum potential.  Here
$R\sim\dot{\mc R}$ is the Ricci curvature associated with the space metric $g_{ij}$
and $\vec{\phi}=-\na log(P)$ is the Weyl vector where $P\sim\hat{\rho}=\rho/\sqrt{|g|}$
for a matter density $\rho$ related to a mass $m$ (note R will then go into the 
SE in a standard manner - cf. \cite{ccz,cczz}).  Then in \cite{cari} we recalled
the flow equation $({\bf 1C})\,\,\pp_tf+\gD f-|\na f|^2+R=0$ 
under Ricci flow with $\int_Mexp(-f)dV=1$.
Independently we showed that $\int_MPQdV\sim{\mf F}$ (where $Q=
-(\hbar^2/2m)(\gD|\psi|/|\psi|)$) with S entering the picture via $({\bf 1D})\,\,\pp_tP+(1/m)
div(P\na S)=0$ leading to $({\bf 1E})\,\,{\mf F}=\int_M(R+|\na f|^2)exp(-f)dV$
(studied in \cite{car}).  Thence via ({\bf 1C}) and $u=exp(-f)=P$ in the form
$({\bf 1F})\,\,P_t=-\gD P+RP$ (Ricci flow situation) we were led to ({\bf 1K}) of \cite{car},
namely $(1/m)div(P\na S)=\gD P-RP$.
The constructions were largely heuristic and preliminary in nature but they suggested a number of possibly interesting connections to eigenvalue problems and lower
bounds on the first eigenvalue of the Laplacian for a Riemannian manifold M (of three
dimensions for convenience here).  One could also consider generalized Ricci flow and 
$\pp_tf=-R-\gD f$ but this seems awkward. 
There are many references for this latter topic and we
mention in particular \cite{aubn,auty,chvl,cheg,heby,hebe,liyu,ling,sace,sacs,tikf,
yau} without specific attribution.  In any event inequalities of the form $({\bf 1G})\,\,\int_M|\na\phi|^2dV\geq\gl_D\int
|\phi|^2dV$ hold for Dirichlet eigenfunctions $\phi$ of the Laplacian.  More generally
for functions in $H^1_0(M)$ with M bounded and $\pp M$ reasonable (e.g. 
weakly convex) one expects $({\bf 1H})\,\,\int_M |\phi|^2dV\leq c\int_M|\na\phi|^2dV$ (cf. \cite{sace})
leading to $\|\phi\|^2_{H_0^1}\leq (1+c)\|\na\phi\|_{L^2}$ as in \cite{car}.  
This in turn leads to solutions of ({\bf 1E}) derived
under the assumption ({\bf 1H}) with M bounded and $0<\gep\leq P(x)$ on M.  Thus
$P\sim |\psi|^2\geq \gep$ on M and $S=0$ on $\pp M$ which seems to involve a 
permissible quantum wave function $\psi=\sqrt{P}exp(iS/\hbar)$.  One is also assuming
some finite time interval (with time treated as a parameter in the S equation).  
\\[3mm]\indent
Another feature of the Ricci flow is developed in \cite{grff} in the form of Ricci
flow gravity.  In \cite{grff} one develops the theory from a geometric Lagrangian
$({\bf 1I})\,\,{\mf L}=\gO(R+\gl(\na\phi)^2)$ in a ``volumetric" manifold $M_4$ where
$\gO=exp(-\phi)\go$ with $\go\sim\sqrt{|g|}\prod_1^4dx^i$.  One notes here that this is precisely
the Lagrangian for conformal general relativity (GR), with no ordinary mass term
(cf. ({\bf 2D})),
and this suggests looking at $M_4\sim M\times T$ with a line element $ds^2=d\gs^2
-dt^2$.  We refer to \cite{ccz,cczz,carl,c3} for a discussion of Brans-Dicke (BD) theory
and conformal GR.  Here $\phi$ in ({\bf 1I}) plays the role of $f$ in ({\bf 1G}) so we can
identify the Perelman functional ${\mf F}$ with some kind of constrained conformal GR action.  Further we showed in \cite{c3} that (in 4-D) this involves an integrable Weyl
geometry with conformal mass $\hat{m}=exp(f/2)m$, which can be identified with a
quantum mass ${\mf M}$ in a corresponding Dirac-Weyl theory, as well as a corresponding
quantum potential $\tl{Q}=(\hbar^2/m^2)(\bx\sqrt{\tl{\rho}}/\sqrt{\tl{\rho}}\,\,(c=1)$ where $|\psi|\sim
\tl{\rho}$ (cf. \cite{ccz,cczz,c3,sh,shg}).  Here ${\mf M}\sim\gb$
where $\gb$ is the Dirac field and ${\mf M}^2/m^2=exp(\tl{Q})$ (cf. \cite{c3,na}).
In \cite{b9} one uses a relativistic quantum potential $\tl{Q}=-(\hbar^2/2m)(\bx|\psi|/|\psi|)$
and we refer to \cite{cczz} for some clarification.

\section{ENHANCEMENT}
\renewcommand{\theequation}{2.\arabic{equation}}
\setcounter{equation}{0}

The theme of Ricci flow gravity in \cite{grff} has a number of attractive features but
we would like to approach it eventually via techniques of Bohmian ``quantum gravity"
following \cite{ccz,cczz,carl,c3,migi,smgi,sh,shg,smg}.  Thus a correct
relativistic quantum equation of motion for a spin zero particle should have the
form $({\bf 2A})\,\,\na^{\mu}S\na_{\mu}S={\mf M}^2c^2$ where $M^2=m^2exp
(\tl{Q})\,\,(\tl{Q}=(\hbar^2/m^2c^2)(\bx|\psi|/|\psi|)$ with $({\bf 2B})\,\,
\na_{\mu}(\rho\na^{mu}S)=0$ where $\rho\sim |\psi|^2$ (see e.g. \cite{cczz,c3,sh}).
Writing $({\bf 2C})\,\,g_{\mu\nu}\to\tl{g}_{\mu\nu}=({\mf M}^2/m^2)g_{\mu\nu}$
(conformal transformation) one arrives at
\bq\label{2.1}
\tl{g}^{\mu\nu}\tl{\na}_{\mu}S\tl{\na}_{\nu}S=m^2c^2;\,\,\tl{g}^{\mu\nu}\tl{\na}_{\mu}
(\rho\tl{\na}_{\nu}S)=0
\end{equation}
This indicates that the presence of a quantum potential is equivalent to having a
curved spacetime with ({\bf 2C}) and can be considered as a geometrization of
the quantum effects of matter.
In order to compare with \cite{grff} we go to \cite{sh,shg,sha} and look at a scalar-tensor
theory with action
\bq\label{2.2}
{\mf A}=
\int\sqrt{-g}d^4x\left[\phi R-\frac{w}{\phi}\na^{\mu}\phi\na_{\mu}\phi+2\gL\phi\right]
\end{equation}
with no ordinary matter Lagrangian.  This is a Brans-Dicke form without ${\mf L}_M$
and is equivalent to conformal GR in the form
\bq\label{2.3}
\hat{{\mf A}}=\int d^4x\sqrt{-\hat{g}}e^{-\psi}\left[\hat{R} -\left(\ga-\frac{3}{2}\right)|\hat{\na}\psi|^2
\right]
\end{equation}
(cf. \cite{c3} - we omit $\gL$ for simplicity and think of $\hat{g}_{ab}=\gO^2g_{ab}$
for $g_{ab}\sim GR$ with $\gO^2=exp(\psi)=\phi=\hat{\phi}^{-1}$ and $[\ga-(3/2)]
\sim w$).  Then $\hat{g}_{ab}\sim ({\mf M}^2/m^2)g_{ab}$ is involved as in ({\bf 2C})
and $({\bf 2D})\,\,{\mf L}=exp(-\psi)\sqrt{-\hat{g}}\prod_1^4dx^i(\hat{R}
-w|\na\psi|^2)$ is exactly the ${\mf L}$ of Graf in ({\bf 1I}) with $\gl\sim -w,\,\,
\psi\sim\phi$, and $\hat{g}\sim g$.
\\[3mm]\indent
{\bf REMARK 2.1.}
The field equations arising from (2.2) (with $\gL=0$) in \cite{sh,shg} are
\bq\label{2.4}
R+\frac{2w}{\phi}\bx\phi+\frac{w}{\phi^2}\na^{\mu}\phi\na_{\mu}\phi=0;\,\,G^{\mu\nu}=
\end{equation}
$$=-\frac{1}{\phi^2}T^{\mu\nu}-\frac{1}{\phi}\left[\na^{\mu}\na^{\nu}-g^{\mu\nu}\bx\right]\phi
+\frac{w}{\phi^2}\na^{\mu}\phi\na^{\nu}\phi-\frac{w}{2\phi^2}g^{\mu\nu}\na^{\ga}\phi
\na_{\ga}\phi$$
where $G_{\mu\nu}=R_{\mu\nu}-(1/2)Rg_{\mu\nu}$ and $T^{\mu\nu}=0$ in the absence of
ordinary matter.
However if we want a quantum contribution in this situation the matter must be introduced and
in \cite{sh,shg} this is done by assuming a matter Lagrangian $({\bf 2E})\,\,L_M=(\rho/m)
\na_{\mu}S\na^{\mu}S-\rho m$ with interaction between the scalar field $\phi$ and the matter
field expressed via $({\bf 2F})\,\,L'_M=(\rho/m)\phi^{\ga}\na^{\mu}S\na_{\mu}S-m\rho\phi^{\gb}
-\gL(1+Q)^{\gag}$ (with suitable choice of $\ga,\,\,\gb,\,\,\gag$ and $1+Q\sim exp(Q)$); this leads to a theory
incorporating both quantum and gravitational effects of matter.  However one can apparently examine quantum effects directly via conformal transformation with $\gO^2=
{\mf M}^2/m^2=exp(\tl{Q})$ and the use of Weyl-Dirac theory with 
the natural Dirac field $\gb\sim{\mf M}$
(cf. \cite{ccz,cczz,c3,sh}).$\hfill\bs$
\\[3mm]\indent
Thus we recall that conformal GR (without $L_M$) is basically equivalent to Weyl-Dirac theory as shown in
\cite{c3} with conformal mass $\hat{m}$ corresponding to $\gb={\mf M}$ where ${\mf M}$
is in turn generated by a ``classical" mass $m$ with ${\mf M}^2=m^2exp(\tl{Q})$ and
$\tl{Q}=(\hbar^2/m^2c^2)(\bx\sqrt{\rho}/\sqrt{\rho})$.  In fact this could serve as a definition
of $\rho$ via ($\sqrt{\rho}=\chi$)
\bq\label{2.5}
\frac{{\mf M}^2}{m^2}=exp(\tl{Q})\Rightarrow \tl{Q}=2log({\mf M}/m)=
\end{equation}
$$=\frac{\hbar^2}{m^2c^2}\frac
{\bx\chi}{\chi}\Rightarrow 2\chi log(\gb/m)=\frac{\hbar^2}{m^2c^2}\bx(\chi)$$
and one could speculate that this approach allow one to geometrize mass in
treating it only via its quantum effect.  However it seems more natural to include 
a matter Lagrangian $L_M$ as well.
Recall also the Weyl vector $w_{\mu}=2\pp_{\mu}log(\gb)$ and in fact (cf. \cite{c3})
\bq\label{2.6}
\hat{{\mf A}}=\int d^4x\sqrt{-\hat{g}}e^{-\tl{Q}}\left[\hat{R}-\left(\ga-\frac{3}{2}\right)(\hat{\na}\tl{Q})^2
+16\pi e^{-\tl{Q}}L_M\right]
\end{equation}
This couples the ordinary and quantum effects of matter directly and gives $\tl{Q}$ a role
as a scalar field (cf. \cite{sh} where Q is made dynamical in a different manner).
\\[3mm]\indent
Now in \cite{grff} one computes from ({\bf 1I}) the terms $({\bf 2G})\,\,(\gd{\mf L}/\gd g_{ik})
\sim P_{ik}$ with $(\gd{\mf L}/\gd\phi)\sim{\mc Q}$ and writes a N\"other identity
\bq\label{2.7}
div(\tl{P}^i_k\xi^k)=P^{ik}L_{\xi}g_{ik}+{\mc Q}L_{\xi}\phi
\end{equation}
followed by a discussion of various situations.  Such an approach is developed systematically
for variational frameworks in e.g. \cite{crpi,fati,gmsi,gsmi,ksmk,sard,saun}
and some concrete examples are given in \cite{cprr,migi} for FRW spaces.  The FRW
framework is very helpful in general and we mention a few details following \cite{akrc,chla,fara,grvk}.  Thus in general one has a metric $({\bf 2H})\,\,ds^2=-dt^2+a^2(t)
\gag_{ij}dx^idx^j$ with $\gag_{ij}=[dr^2/(1-kr^2)]+r^2d\gO^2_2$ in 4-D, where $d\gO^2_2$
is the metric for a 2-D sphere with unit radius.  If $R_{ij}$ is the Ricci tensor for $\gag_{ij}$
then $({\bf 2I})\,\,R_{ij}=2k\gag_{ij}$ and $k=1,0,-1$ corresponding to positive, zero, or
negative curvature for the 3-D hypersurface (note $d\gO^2_2\sim d\gt^2+Sin^2(\gt)d\phi^2$).
\\[3mm]\indent
The Jordan-Brans-Dicke action (with $\gL=0$) is then given by (2.2) with $L_M=
\int d^4x\sqrt{-g}L_M$ added and the equations of motion can be developed as follows
(cf. \cite{chla,fara}).  First $\phi$ is to depend only on $t$ so $\na^c\phi\na_c\phi=
-(\dot{\phi})^2$ and 
\bq\label{2.8}
\bx\phi=-(\ddot{\phi}+3H\dot{\phi})=-\frac{1}{a^3}\frac{d}{dt}(a^3\dot{\phi})
\end{equation}
where $H=(\dot{a}/a)$ is the Hubble parameter.  Further
\bq\label{2.9}
T_{ab}=-\frac{2}{\sqrt{-g}}\frac{\gd}{\gd g^{ab}}(\sqrt{-g}L_M)=(P+\rho)u_au_b+Pg_{ab};
\,\,T=3P-\rho
\end{equation}
where $\rho=$ energy density and P is the pressure of a ``cosmic fluid" moving with
velocity $u^{\mu}$.  The field equations are
\bq\label{2.10}
G_{ab}=\frac{8\pi}{\phi}T_{ab}+\frac{w}{\phi^2}\left(\na_a\phi\na_b\phi-\frac{1}{2}g_{ab}
\na^c\phi\na_c\phi\right)+
\end{equation}
$$+\frac{1}{\phi}(\na_a\phi\na_b\phi-g_{ab}\bx\phi)-\frac{V}{2\phi}g_{ab}$$
and 
\bq\label{2.11}
\frac{2w}{\phi}\bx\phi+R-\frac{w}{\phi^2}\na^c\phi\na_c\phi-\frac{dV}{d\phi}=0
\end{equation}
One obtains then from (2.10)
\bq\label{2.12}
R=-\frac{8\pi T}{\phi}+\frac{w}{\phi^2}\na^c\phi\na_c\phi+\frac{3\bx\phi}{\phi}+\frac{2V}{\phi}
\end{equation}
which implies (using $R=6[\dot{H}+2H^2+(k/a^2)]$)
\bq\label{2.13}
\dot{H}+2H^2+\frac{k}{a^2}=-\frac{4\pi T}{3\phi}-\frac{w}{6}\left(\frac{\dot{\phi}}{\phi}\right)^2
+\frac{1}{2}\frac{\bx\phi}{\phi}+\frac{V}{3\phi}
\end{equation}
Further from (2.10) one has
\bq\label{2.14}
H^2=\frac{8\pi}{3\phi}\rho+\frac{w}{6}\left(\frac{\dot{\phi}}{\phi}\right)^2-H\frac{\dot{\phi}}
{\phi}-\frac{k}{a^2}+\frac{V}{6\phi}
\end{equation}
which provides a first integral.  Further from (2.11) and (2.12) follows also
\bq\label{2.15}
\bx\phi=\frac{1}{2w+3}\left[8\pi T+\phi\frac{dV}{d\phi}-2V\right]
\end{equation}
and putting together (2.14) and (2.15) with $T=3P-\rho$ yields then
\bq\label{2.16}
\dot{H}=\frac{-8\pi}{(2w+3)\phi}[(w+2)\rho+wP]-\frac{w}{2}\left(\frac{\dot{\phi}}{\phi}\right)^2+
\end{equation}
$$+2H\frac{\dot{\phi}}{\phi}+\frac{k}{a^2}+\frac{1}{2(2+3)\phi}\left(\phi\frac{dV}{d\phi}-2V\right)$$
Moreover from (2.15)
\bq\label{2.17}
\ddot{\phi}+3H\dot{\phi}=\frac{1}{2w+3}\left[8\pi(\rho-3P)-\phi\frac{dV}{d\phi}+2V\right]
\end{equation}
A remark might be appropriate here.
\\[3mm]\indent
{\bf REMARK 2.2.}
In \cite{chla} one uses units with $8\pi G=1=c=\hbar$ and instead of a standard $L_M$
an inflaton field Lagrangian $L(\psi)=(1/2)\pp_{\mu}\psi\pp^{\mu}\psi-U(\psi)$ is used
(also $k=1$ is assumed).  This leads to a conservation equation $({\bf 2J})\,\,
\dot{\rho}+3H(\rho+P)=0\equiv \ddot{\psi}+3H\dot{\psi}=-\pp_{\psi}U$ with $({\bf 2K})\,\,
\rho=(1/2)\dot{\psi}^2+U(\psi)$ and $P=(1/2)\dot{\psi}^2-U(\psi)$.$\hfill\bs$
\\[3mm]\indent
What we want now is the Ricci curvature determined by the space metric 
$\gag_{ij}$ as in ({\bf 2H}) and we refer to \cite{call,glka,rbts} for input. 
In \cite{call} for example one uses a line element
\bq\label{2.18}
ds^2=-dt^2+a^2(t)\left[\frac{dr^2}{1-\gk r^2}+r^2d\gO^2\right]
\end{equation}
Here $\gk$ can take any value but it is related to $(+,0,-)$ curvatures according to sign.
One sets $\dot{a}=da/dt$ and computes Christoffel symbols to be
\bq\label{2.19}
\gG^0_{11}=\frac{a\dot{a}}{1-\gk r^2};\,\,\gG^1_{11}=\frac{\gk r}{1-\gk r^2};\,\,\gG^0_{22}= 
a\dot{a}r^2;\,\,\gG^0_{33}=a\dot{a}r^2Sin^2(\gt);
\end{equation}
$$\gG^1_{01}=\gG^2_{02};\,\,\gG^3_{03}=\frac{\dot{a}}{a};\,\,\gG^1_{22}=-r(1-\gk r^2);\,\,
\gG^1_{33}=-r(1-\gk r^2)Sin^2(\gt);$$
$$\gG^2_{12}=\gG^3_{13}=\frac{1}{r};\,\,\gG^2_{33}
=-Sin(\gt)Cos(\gt);\,\,\gG^3_{23}=Ctn(\gt)$$
leading to
\bq\label{2.20}
R_{00}=-3\frac{\ddot{a}}{a};\,\,R_{11}=\frac{a\ddot{a}+2\dot{a}^2+2\gk}{1-\gk r^2};
\end{equation}
$$R_{22}=r^2[a\ddot{a}+2\dot{a}^2+2\gk];\,\,R_{33}=r^2[a\ddot{a}+
2\dot{a}^2+2\gk]Sin^2(\gt)$$
with Ricci scalar 
\bq\label{2.21}
{}^4R=6\left[\frac{\ddot{a}}{a}+\left(\frac{\dot{a}}{a}\right)^2+\frac{\gk}{a^2}\right]=
{}^3R+3\frac{\ddot{a}}{a}
\end{equation}
Thus if we write now $g_{00}=-1$ with $\tl{g}_{ii}$ given via (2.19) we consider Ricci flow
for the space metric $\tl{g}_{ij}=a^2(t)\gag_{ii}$ in the form $\pp_t\tl{g}_{ii}=-2\,{}^3R_{ii}\,\,
(i=1,2,3)$ to arrive at $({\bf 2L})\,\,2a\dot{a}\gag_{ii}=-2{}^3R_{ii}$.  Since $\gag^{ii}\gag_{ii}
=3$ we obtain $({\bf 2M})\,\,6a\dot{a}=-2\,{}^3R$ which can be written as
\bq\label{2.22}
6a\dot{a}=-2\left[3\frac{\ddot{a}}{a}+6\left(\frac{\dot{a}}{a}\right)^2+\frac{6\gk}{a^2}\right]
\end{equation}
\begin{proposition}
Ricci flow in the form $\pp_t\tl{g}_{ii}=-2\,{}^3R_{ii}$ with $\tl{g}_{ij}=a^2(t)\gag_{ij}$ as in 
(2.18) leads to (2.22) as a stipulation about $a(t)$.
\end{proposition}
\indent
Now to involve the Einstein equations $R_{\mu\nu}=8\pi G(T_{\mu\nu}-(1/2)g_{\mu\nu}T)$
the $\mu\nu=00$ equation gives $({\bf 2N})\,\,-3(\ddot{a}/a)=4\pi G(\rho+3P)$ and the
$\mu\nu$ equations give
\bq\label{2.23}
\frac{\ddot{a}}{a}+2\left(\frac{\dot{a}}{a}\right)^2+2\frac{\gk}{a^2}=4\pi G(\rho-P)
\end{equation}
Due to isotropy there is only one distinct equation from $\mu\nu=ij$ and one obtains
\bq\label{2.24}
\left(\frac{\dot{a}}{a}\right)^2=\frac{8\pi G}{3}\rho-\frac{\gk}{a^2};\,\,\frac{\ddot{a}}{a}
=-\frac{4\pi G}{3}(\rho+3P)
\end{equation}
which are known as the Friedman equations (note $P<-(1/3)\rho$ implies repulsive
gravitation).  Recalling the Hubble parameter $H=\dot{a}/a$ these can be written as
\bq\label{2.25}
H^2=\frac{8\pi G}{3}\rho-\frac{\gk}{a^2};\,\,\dot{H}=-4\pi G(\rho+P)+\frac{\gk}{a^2}
\end{equation}
We will discuss (2.22) and (2.24) in Section 3.
\\[3mm]\indent
{\bf REMARK 2.3.}
From \cite{grvk} we note that for a universe in a state of accelerated expansion ($\ddot{a}
>0$) the second Friedman equation in (2.24) implies $\rho+3P<0$ and for a vacuum
$P=-\rho$ so $\rho+3P=-2\rho<0$ (repulsive gravity).  The Friedman
equations with a cosmological constant for a homogeneous isotropic universe models
with pressure free matter are
\bq\label{2.26}
\frac{\ddot{a}}{a}=\frac{\gL}{3}-\frac{4\pi G}{3}\rho;\,\,H^2=\left(\frac{\dot{a}}{a}\right)^2
=\frac{\gL}{3}-\frac{k}{a^2}+\frac{8\pi G}{3}\rho
\end{equation}
(see Section 3 for more on this).$\hfill\bs$
\\[3mm]\indent
{\bf REMARK 2.4.}
In \cite{kief}, p. 210, one has $ds^2=-N^2(t)dt^2+a^2(t)d\gO^2_3$ with $d\gO^2_3=
d\chi^2+Sin^2(\chi)(d\gt^2+Sin^2(\gt)d\phi^2)$, second fundamental form $({\bf 2O})
\,\,K_{ab}=-(1/2N)(\pp h_{ab}/\pp t)=-(\dot{a}/aN)h_{ab}$, and $K=K_{ab}h^{ab}=
-3\dot{a}/aN$ (cf. also \cite{glka}).  Given the EH action
\bq\label{2.27}
S_{EH}=\frac{c^4}{16\pi G}\int_Md^4x\sqrt{-g}(R-2\gL)\pm\frac{c^4}{8\pi G}\int_{\pp M}
d^3x\sqrt{h}K
\end{equation}
we obtain for the surface term $({\bf 2P})\,\,2\int d^3xK\sqrt{h}=-6\int d^3x(\dot{a}/Na)\sqrt{h}$
with $\sqrt{h}=a^3Sin^2(\chi)Sin(\gt)$.  This surface term is cancelled after partial
integration and one obtains (setting $2G/3\pi=1$)
\bq\label{2.28}
S_g=\frac{1}{2}\int dt N\left(-\frac{a\dot{a}^2}{N^2}+a-\frac{\gL a^3}{3}\right)
\end{equation}
The matter action is given by (after a rescaling $\phi\to \phi/\sqrt{2}\pi$) a formula
$({\bf 2Q})\,\,S_m=(1/2)\int dt Na^3[(\dot{\phi}^2/N^2)-m^2\phi]$ leading to a minisuperspace
action $(q^1\sim a$ and $q^2\sim \phi$)
\bq\label{2.29}
S=S_q+S_m=\int dt N\left[\frac{1}{2}G_{AB}\frac{\dot{q}^A\dot{q}^B}{N^2}-V(q)\right];\,\,
G_{AB}=\left(\begin{array}{cc}
-a & 0\\
0 & q^3
\end{array}\right)
\end{equation}
with $\sqrt{-G}=a^2$.
Following general techniques (cf. \cite{kief}) this leads to a Wheeler-deWitt (WDW) equation
\bq\label{2.30}
\frac{1}{2}\left(\frac{\hbar^2}{a^2}\pp_a(a\pp_a)-\frac{\hbar^2}{a^3}\frac{\pp^2}{\pp
\phi^2}-a+\frac{\gL a^3}{3}+m^2a^3\phi^2\right)\psi(a,\phi)=0
\end{equation}
and this has a simpler form upon writing $\ga=log(a)$, namely
\bq\label{2.31}
\frac{e^{-3\ga}}{2}\left(\hbar^2\frac{\pp^2}{\pp\ga^2}-\hbar^2\frac{\pp^2}{\pp\phi^2}-
e^{4\ga}+e^{6\ga}\left[m^2\phi^2+\frac{\gL}{3}\right]\right)\psi=0
\end{equation}
This has the form of a Klein-Gordon equation with potential $V(\ga,\phi)=-exp(4\ga)+
exp(6\ga)[m^2\phi^2+(\gL/3)]$.$\hfill\bs$

\section{SOME HEURISTIC RESULTS}
\renewcommand{\theequation}{3.\arabic{equation}}
\setcounter{equation}{0}

For general background we refer to \cite{call,ciwh,grvk,hals,rind,wald,wein} and
for Bohmian aspects see \cite{ccz,cczz,carl,c3,migi,na,smgi,sh,shg,smg,sha}.
Some of this is mentioned in Section 2 and we add here first a few comments.
\\[3mm]\indent
The two most popular examples of cosmological fluids are known as matter and radiation.
Matter involves collisionless, nonrelativistic particles having essentially zero pressume
$P_M=0$ (e.g. stars and galaxies or dust); the energy density in matter falls off as
$({\bf 3A})\,\,\rho_M\propto a^{-3}$.
For electromagnetic energy one has $({\bf 3B})\,\,T^{\mu\nu}=F^{\mu\gl}F^{\nu}_{\gl}-(1/4)g^{\mu\nu}F^{\gl\gs}F_{\gl\gs}$ with trace 0; since this must equal $T=-\rho+3P$ one has
$P_R=(1/3)\rho_R$.  A universe dominated by radiation is then postulated to have
$({\bf 3C})\,\,\rho_R\propto a^{-4}$; for a perfect fluid with $p_{\gL}=-\rho_{\gL}$ one assumes
$\rho_{\gL}\propto a^0$ (vacuum dominated).
\\[3mm]\indent
There is now a density parameter $({\bf 3D})\,\,\gO=(8\pi G/3H^2)\rho=\rho/\rho_{crit}$ and from (2.24) one has $({\bf 3E})\,\,\gO-1=\gk/H^2a^2$.  Thus the sign of $\gk$ is determined
via
\begin{enumerate}
\item
$\rho<\rho_{crit}\sim \gO<1\sim\gk<0\,\,(open)$
\item
$\rho=\rho_{crit}\sim\gO=1\sim \gk=0\,\,(flat)$
\item
$\rho>\rho_{crit}\sim\gO>1\sim \gk>0\,\,(closed)$
\end{enumerate}
We refer to \cite{call} for more discussion.
\\[3mm]\indent
{\bf REMARK 3.1.}
We gather here a few observations from \cite{grvk}.  Note first from the Friedman equations
that one can write $({\bf 3F})\,\,\dot{\rho}+3(\dot{a}/a)(\rho+P)=0\equiv (d/dt)(\rho a^3)+
P(d/dt)a^3=0$.  Let then $V\sim a^3$ correspond to the volume of an expanding region together with the ``cosmic fluid" (a comoving volume).  The energy is then $U=\rho a^2$ so that
({\bf 3F}) becomes $dU+PdV=0$.  The first law of thermodynamics states that $({\bf 3G})\,\,
TdS=dU+PdV$ for a fluid in equilibrium and here $dS=0$ corresponds to an adiabatic
process.  If in addition $P=w\rho$ one obtains from ({\bf 3F}) the relation $({\bf 3H})\,\,
(d/dt)(\rho a^3)+w\rho(d/dt)a^3=0\Rightarrow \rho a^{3(w+1)}=\rho_0$. 
Lorentz invariant vacuum energy (LIVE) involves $w=-1$ whereas for dust $w=0$ and 
radiation involves $w=1/3$.  For $P=w\rho$ the first Friedman equation becomes
\bq\label{3.1}
\left(\frac{\dot{a}}{a}\right)^2=\frac{8\pi G}{3}\frac{\rho_0}{a^{3(1+w)}}-\frac{\gk}{a^2}
\end{equation}
and we refer to \cite{grvk} for a discussion of phenomena depending on $w>-(1/3),\,\,
w<-(1/3)$, or $w=-1/3$ (as well as many other matters).$\hfill\bs$
\\[3mm]\indent
{\bf REMARK 3.2.}
From \cite{grvk} again the Einstein equations with a cosmological constant are
\bq\label{3.2}
R_{\mu\nu}-\frac{1}{2}g_{\mu\nu}R+\gL g_{\mu\nu}=8\pi GT_{\mu\nu}
\end{equation}
and for a vacuum ($T_{\mu\nu}=0$) a homogeneous isotropic model with $\gL>0$
involves equations
\bq\label{3.3}
3\frac{\dot{a}^2+k}{a^2}-\gL=0;\,\,-2\frac{\ddot{a}}{a}-\frac{\dot{a}^2+k}{a^2}+\gL=0
\end{equation}
(this is just (2.26) rewritten for $\rho=0$).  The first equation is $\dot{a}^2-\go^2a^2=
-k$ with solutions
\bq\label{3.4}
a(t)=\left\{\begin{array}{cc}
\frac{\sqrt{k}}{\go}Cosh(\go t) & (k>0)\\
exp(\go t) & (k=0)\\
\frac{\sqrt{|k|}}{\go}Sinh(\go t) & (k<0)
\end{array}\right.
\end{equation}
This gives a deSitter model and it is instructive to draw graphs for (3.20).  For $k=0$ one
has $H=(\dot{a}/a)=\go$ and $a^2=exp(2Ht)$ with $ds^2=-dt^2+exp(2Ht)(dx^2+dy^2+dz^2)$.
We refer to \cite{grvk} for further discussion about event and particle horizons (cf. also
\cite{ciwh,hals,rind} for information about such horizons).$\hfill\bs$
\\[3mm]\indent
Consider now Ricci flow via (2.22) in conjunction with the Friedman equations
(2.24).  One obtains first
\bq\label{3.5}
3\frac{\ddot{a}}{a}+6\left(\frac{\dot{a}}{a}\right)^2=-4\pi G(\rho+3P)+6\left[\frac
{8\pi G}{3}\rho-\frac{\gk}{a^2}\right]=12\pi G(\rho-P)-\frac{6\gk}{a^2}
\end{equation}
which means that
\begin{proposition}
Ricci flow plus the Friedman equations (2.24) implies
\bq\label{3.6}
6a\dot{a}=-2[12\pi G(\rho-P)]\Rightarrow \frac{d}{dt}a^2=8\pi G(P-\rho)
\end{equation}
(independently of $\gk$).
\end{proposition}

\indent
{\bf REMARK 3.3.}
We assume here that the Ricci flow persists in time, ignoring much of the 
Hamilton-Perelman analysis (cf. \cite{chln,khol,khlb,muhl,perl,topp}).  Then 
(3.6) says that $a^2$ is increasing (resp. decreasing) for $P>\rho$ (resp. $P<\rho$).
This gives a somewhat different perspective in comparison with that arising from
the Friedman equations (2.24).  We note that (2.24) (or (2.26)) do not specify a sign
for $\dot{a}$.  In particular for $P=0$ or $P=-\rho$ (3.6) implies that $\dot{a}<0$
for $\rho>0$.$\hfill\bs$
\\[3mm]\indent
{\bf REMARK 3.4.}
From (2.25) (resp. (2.26)) in order to have $H^2\geq 0$ one requires
\bq\label{3.7}
\frac{8\pi G}{3}\rho\geq\frac{\gk}{a^2}\,\,or\,\,\frac{8\pi G}{3}+\frac{\gL}{3}\geq
\frac{\gk}{a^2}
\end{equation}
We have then from (2.26) and (3.7B)
\bq\label{3.8}
2\frac{\ddot{a}}{a}=\frac{2\gL}{3}-\frac{8\pi G}{3}\rho\geq \frac{2\gL}{3}-\left(
\frac{\gk}{a^2}-\frac{\gL}{3}\right)\geq\gL-\frac{\gk}{a^2}
\end{equation}
which seems to predict repulsive gravity for $\gL-(\gk/a^2)>0$.$\hfill\bs$
\\[3mm]\indent
One can of course ask for an existence (uniqueness) theorem for the ordinary
differential equation (ODE) (2.22) on some time interval.  This can be rewritten in
terms of H via $\dot{H}=(\ddot{a}/a)-H^2$ to obtain
\bq\label{3.9}
\dot{H}+3H^2+\frac{2\gk}{a^2}-a^2H=0
\end{equation}
Set then $\chi=a^3$ and $\tau=3t$ so that $H=\chi_{\tau}/\chi$ (note $\chi_{\tau}=
(1/3)\dot{\chi}=(1/3)3a\dot{a}$ and $\chi_{\tau}/\tau=(a^2\dot{a}/a^3)=H$).  Then
(3.9) becomes a Ricatti type equation (note $\dot{H}=3H_{\tau}$)
\bq\label{3.10}
H_{\tau}+H^2+\frac{2\gk}{3a^2}-\frac{a^2}{3}H=0\equiv \chi_{\tau\tau}-
\frac{a^2}{3}\chi_{\tau}+\frac{2\gk}{3a^2}\chi=0
\end{equation}
Consequently 
\bq\label{3.11}
\left[\chi_{\tau}e^{-\int_0^{\tau}(a^2/3)ds}\right]_{\tau}+\frac{2\gk}{3a^2}\chi=0
\Rightarrow \chi_{\tau}=e^{-\int_0^{\tau}ds}\left[c+\frac{2\gk}{3}\int_0^{\tau}
\frac{\chi}{a^2}ds\right]
\end{equation}
($c\sim \chi_{\tau}(0)$) leading to
\bq\label{3.12}
2a^2a_{\tau}=e^{-\int_0^{\tau}(a^2/3)ds}\left[c+\frac{2\gk}{3}\int_0^{\tau}ads\right]
\end{equation}
and consequently
\bq\label{3.13}
a^3=a_0^3+\int_0^{\tau}e^{-\int_0^{\ga}(a^2/3)ds}\left[c+\frac{2\gk}{3}\int_0^{\ga}
ads\right]d\ga
\end{equation}
\begin{proposition}
Equation (3.12) is an ODE of the form $({\bf 3I})\,\,a_{\tau}=J(a,\tau)$ with J 
Lipschitz in any region $0<\gep\leq a\leq A$ (and say $0\leq \tau\leq T$).  Hence
via standard theorems (see e.g. \cite{caro}) there is a unique local solution.
\end{proposition}

\indent
There is then a question of the behavior of $a$ as a solution of the integrodifferential
equation (3.13) or the ODE (3.12).  This will depend in particular on the signs of $c$
and $\gk$ and we leave this for a subsequent article.

\end{document}